\documentclass[10pt,b5paper,twoside]{myarticle}

\usepackage{graphicx}
\usepackage{natbib}

\begin{document}

\title{The radio colour--colour diagram of Van der Laan Bubbles 
--- an application to SS433}
\author{Zsolt Paragi\thanks{This work has been carried out in collaboration 
with I. Fejes, R.C. Vermeulen, R.T. Schilizzi, R.E. Spencer and A.M. Stirling. Special thanks to Al Stirling for useful suggestions, Istv\'an Fejes and 
S\'andor Frey for careful reading. Financial support is acknowledged
from the Hungarian Space Office (M\H{U}I), the Netherlands Organization for
Scientific Research (NWO) and the Hungarian Scientific Research Fund (OTKA,
grant no. N31721 \& T031723).} \\
\institutename{Joint Institute for VLBI in Europe / F\"OMI Satellite Geodetic Observatory } \\
\instituteaddress{Postbus 2, 7990 AA Dwingeloo, The Netherlands} \\
\email{paragi@jive.nl}}

\date{}

\maketitle

\begin{abstract}
\noindent

A radio "colour--colour" diagram is defined in order to determine the
evolutionary state of synchrotron-emitting plasma bubbles that
are ejected during outbursts in microquasars. We establish the 
colour--colour diagram for the plasmons of SS433 observed on 18 April 
1998 using VLBI observations. We show that the radio plasmons are not
consistent with a simple expanding sphere of plasmon. This may indicate 
that in-situ particle acceleration is taking place away from the 
central engine.

\end{abstract}

\section{Introduction}

Microquasars are galactic X-ray binary systems that contain a normal star 
and a neutron star or black hole. The former loses matter onto the compact 
object through an accretion disk. Part of this accreted matter leaves the 
system in well collimated particle beams (jets), perpendicular to the disk. 
The nature of the compact object is still not known in several systems. 
How jets are formed, collimated, and how energetic electrons are produced in 
these relativistic beams are open questions. Most of these energetic
electrons probably 
originate from the vicinity of the central engines of microquasars, but the 
Fermi-process (electron acceleration in shocks) within the jets may also play 
a role. Jet processes are briefly summarized by \citet{Spencer98}.

An introduction to the Galactic radio-jet system SS433 as a microquasar and 
its high resolution properties determined in recent VLBI observations is 
given in \citet{Paragi01}. The particular VLBI experiment shown here and 
the data analysis are described in \citet{Paragi00}.

Below we introduce the \citet{VanderLaan66} model (originally developed for
quasars) that describes radio emission from spherical plasmons ejected from 
the central engine during an outburst. The spectral index evolution of these 
ejecta from the optically thick to the optically thin regime is demostrated 
on a radio colour-colour diagram. A spectral analysis of the radio components 
observed in SS433 follows.

\section{Radio emission from spherically symmetric ejecta}

Radiative properties of spherically symmetric ejecta depend on their
instantaneous  apparent size ($\theta$), the number density of relativistic
electrons ($N$), and the strength of the magnetic field ($H$). The energy
distribution of synchrotron
radiating particles is $N(E)=KE^{-s}$, where $s$ is the energy spectral
index. The emission and absorption coefficients of the synchrotron process
have the following dependence on frequency: $j_{\nu}\propto\nu^{(1-s)/2}$ and
$\kappa_\nu\propto\nu^{-(4+s)/2}$, respectively. The resulting spectrum is 
$S\propto\nu^{\alpha}$, $\alpha$ is the spectral index.
In the optically thin domain the emission coefficient determines the spectrum,
and $\alpha=(1-s)/2$. In the optically thick domain the spectrum depends on 
the source function ($j_{\nu}/\kappa_{\nu}$), resulting in $\alpha=2.5$.

The time evolution of the received flux density ($S$) in a radio outburst
as observed at different frequencies 
was calculated by \citet{VanderLaan66}.
The effect of the optical depth ($\tau$) changing through different lines of sights in the source was considered by \citet{HJ88}. Their basic assumptions
were that the ejected cloud of plasma expands adiabatically into the 
surrounding medium, and relativistic electrons are genarated within a short 
time range close to the central engine, and not somewhere in the extended 
the jet. This latter assumption seems 
to be valid for microquasars \citep{Spencer98}. 

\begin{figure}[htb]
\centering
\resizebox{120mm}{!}
{
\includegraphics[angle=-90]{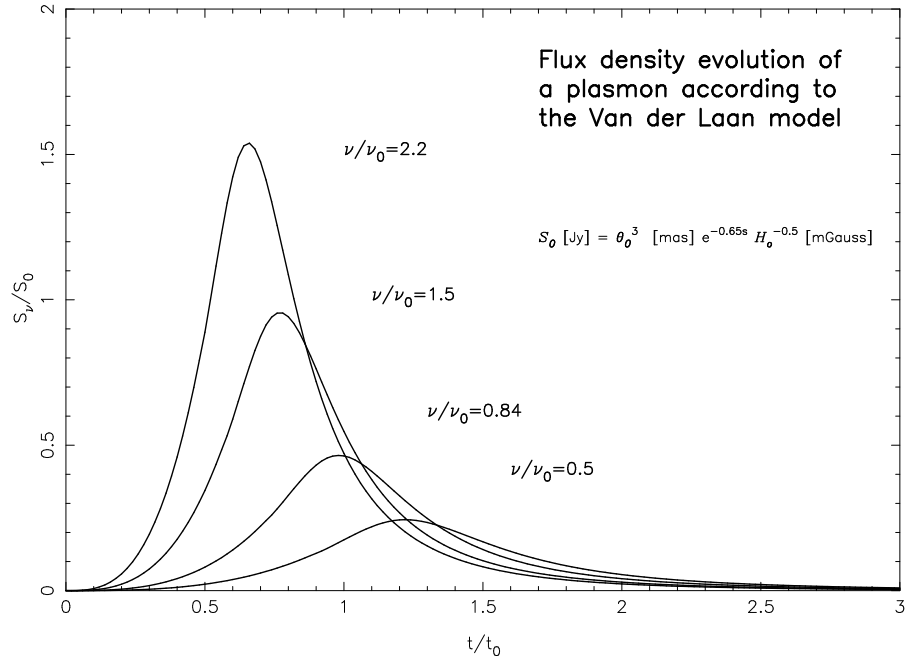}
}
\caption{Flux density evolution of plasmons \citep{VanderLaan66,HJ88}.
Observing frequencies available in VLBI were selected 
($\nu_{0}=$10 GHz, $\nu=$22, 15, 8.4, 5 GHz)}
\label{fig:fluxdens}
\end{figure}

The time evolution of the flux density at four different frequencies is shown 
in Fig.~\ref{fig:fluxdens}. The plasmon brightens in the optically thick stage
(until $\tau\sim 1$) with increasing radius ($\theta\propto t$ for free
expansion), and there is an exponential cutoff in the optically thin regime.
Of course $\tau$ has a strong frequency dependence, this is why the radio
lightcurves peak at different times.


\section{The radio colour-colour diagram}

The model outlined above might be checked in multi-frequency observations by
monitoring the total flux density of the sources during outburst. However,
there may be several radio components contributing to the total flux, 
therefore it is desirable to have high resolution imaging experiments using 
the Very Long Baseline Interferometry Technique \citep{Zen95}. 
VLBI observations are useful also because we can measure the size of the
ejected plasmons, and this allows us to estimate the magnetic field strength. 
Even more information could be gathered if we could determine the optical 
depth of the component directly. In order to achieve this,  we define a 
radio "colour--colour" diagram.

Similarly to Fig.~\ref{fig:fluxdens}, one might plot the spectral index
evolution with time, starting from $\alpha$=2.5 and eventually reaching
$\alpha$=($1-s$)/2 (not shown). 
In practice we determine the spectral index between two observing
frequencies as $\alpha_{ij}=\rm{ln}(S_{i}/S_{j})/\rm{ln}(\nu_{i}/\nu_{j}))$
($\nu_{i}>\nu_{j}$). It is straightforward that the high frequency spectral
index versus the low frequency spectral index plot (the "colour--colour" diagram) will be model dependent, and contains information about the evolutionary state
(i.e. the optical depth) of plasmons at these frequencies. The radio 
colour--colour diagram of a model outburst is compared to real measurements 
of SS433 in the next section.

\section{Application to SS433}

\begin{figure}[htb]
\centering
\resizebox{120mm}{!}
{
\includegraphics[angle=-90]{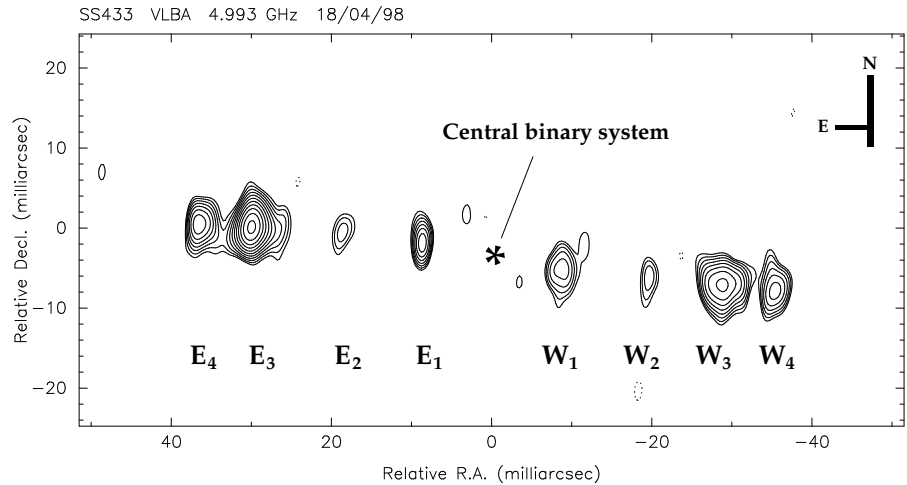}
}
\caption{The plasmons of SS433 during a flare on 18 April 1998} 
\label{fig:map}
\end{figure}

The source was observed at four frequencies on 18 April 1998, during 
a large flare. The VLBI map of SS433 at 5 GHz is shown in 
Fig.~ \ref{fig:map}. There are four pairs of radio components, their ages 
range between $\sim 1-4$ days. We have the opportunity to observe 
plasmons of different age, i.e. in different evolutionary states even though 
the observations were made at a single epoch.

\begin{figure}[htb]
\centering
\resizebox{120mm}{!}
{
\includegraphics[angle=-90]{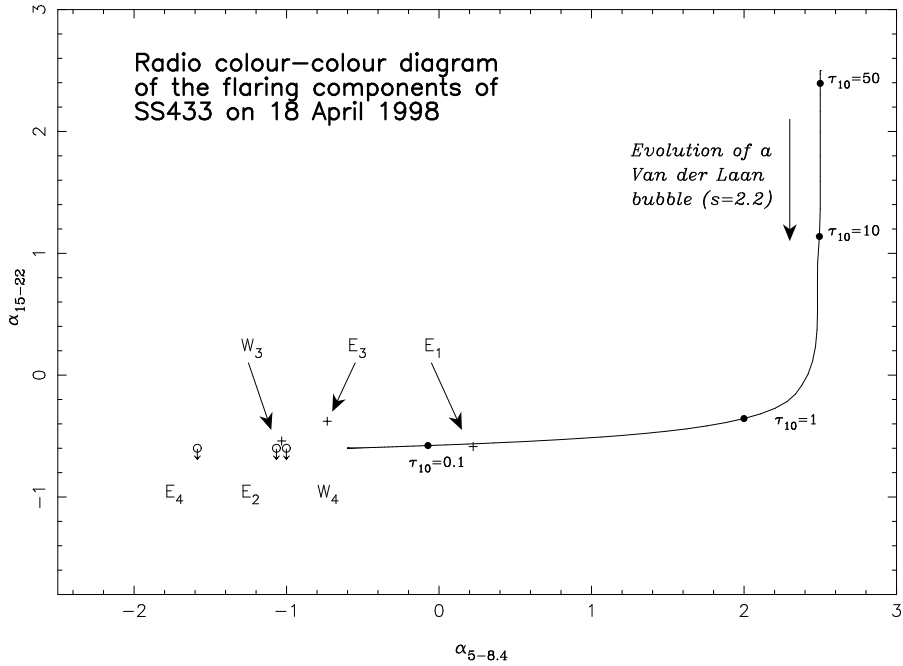}
}
\caption{The SS433 radio components compared to the spherical plasmon model
(solid curve).
There are upper limits for $\alpha_{15-22}$ of some of the components that 
were not detected at the highest frequencies, W$_{2}$ was too faint even at
8~GHz, and W$_{1}$ lies out the ranges shown. The energy spectral index used 
in the model is $s=2.2$}
\label{fig:plot}
\end{figure}

It can be seen from the radio colour--colour diagram (Fig.~\ref{fig:plot}) 
that the plasmons of SS433 
are already in the optically thin regime at these frequencies. Even the 
youngest component has $\tau_{10}\sim 0.1$. Other components have spectral
indices that are not compatible with the model (W$_{1}$ is located
outside the ranges shown). It seems that SS433 outbursts cannot be explained 
by spherically symmetric plasmons as described by \citet{VanderLaan66}
-- one or more model assumptions must be invalid. 

One may speculate that the radio components seen are not spherical bubbles 
but shocked regions within the jet. In this case we expect to see significant
degrees of linearly polarized emission, because shocks are ordering the 
magnetic field and the jets are near the plane of the sky (also, aberration
is unimportant at the jet speed of 0.26c). We do not detect polarization in
SS433 on VLBI scales. Unless there is an external Faraday screen depolarizing
the source (to be investigated later), this means that the radio components 
seen on the maps are not single shocks.

There is one model assumption identified so far that is surely not true. 
We have shown that the components are optically thin. But we are in the 
midst of a large flare in integrated flux density, and $S$ is expected 
to increase only 
in the optically thick regime. In order to increase the brightness of an
optically thin source, relativistic electrons must be continuously produced 
-- these particles are not supplied by the central engine in this case!
Energetic electrons may be accelerated via the Fermi-process in shocks, 
but these shocks must be smaller than the components seen on the maps
for two reasons: {\it i)} we need several
crossings through the shock in order to achieve an efficient acceleration 
(the smaller the shock the more effective the process is), and {\it ii)} 
small shocked regions with different orientations of $H$ may cancel out 
the net polarization, resulting in a depolarized source as we observe.

\section{Conclusions}

We define a radio colour--colour diagram and demonstrate its applicability 
in analysing radio flares by comparing a model with the observations. 
There must be ongoing production of relativistic particles in order to 
explain the optically thin nature of the components and at the same time 
the increasing brightness of the source. This is clear evidence that in
microquasars the central engine is not the only place where electrons can 
be accelerated to relativistic energies.


\end{document}